\documentstyle[12pt]{article}

\newcommand{\be}{\begin{equation}}
\newcommand{\ee}{\end{equation}}

\newcommand{\bea}{\begin{eqnarray}}
\newcommand{\eea}{\end{eqnarray}}

\newcommand{\p}{\partial}

\newcommand{\nn}{\nonumber \\}
\newcommand{\f}{\frac}
\newcommand{\w}{\wedge}

\begin{document}

\thispagestyle{empty}

\begin{flushright}
arXiv:0808.3042
\end{flushright}
\begin{center} \noindent \Large \bf
 Null Melvin Twist to Sakai-Sugimoto model
\end{center}

\vskip 0.5cm
\begin{center}
{ \normalsize {\bf  Shesansu Sekhar Pal}\\
\vskip 0.5cm

 Barchana, Jajpur, 754081, Orissa, India \\
\vskip 0.5 cm

 S N Bose National Centre for Basic Sciences,\\
 Sector III, Block JD, Salt lake, Kolkata 700098, India \\
\vskip 0.5 cm

\sf shesansu${\frame{\shortstack{AT}}}$gmail.com}
\end{center}

\centerline{\bf \small Abstract}
The   Null Melvin Twist to D4 brane background  has been studied
 and putting in a bunch of coincident 
D8 and anti-D8 branes {\it a la} Sakai-Sugimoto  model results in a 
 breakdown of a global symmetry to its diagonal subgroup. 
The same is studied at finite temperature which 
results in the unchanged form of the chiral symmetry breaking and restoration
curve. The effect of turning on a finite chemical potential do not changes 
the structure of the phase diagram with respect to Null Melvin Twist.
 Moreover, the confining and de-confining 
phase transition has the same structure as in the absence of  
Null Melvin Twist.

\newpage
\section{Introduction}

It is certainly important and interesting to understand the strongly coupled
feature of any  field theory but in practice it is not an easy job. 
However, it can be made easy, in some limit, if we know the corresponding 
gravity dual. In this context, if we know a system that
preserve some symmetry may be relativistic or non-relativistic type, then 
the way to 
understand is by the prescription of gauge gravity correspondence, which means 
one need to find an appropriate  geometric description for a system whose 
dual field theory at strong coupling exhibiting the above  symmetry.

In general it is not easy to find such a gravitational description, but given
a gravity description which  respects that symmetry, it is easy to generate
another  gravitational description, using the symmetry of the problem.
In this context, if we start with a theory that has, let us say,
 Poincare symmetry then
using the  solution generating techniques we can have new solutions  
for which the system can have the same or different symmetry. Here we are 
interested in the subgroup of the original symmetry group i.e. it 
possesses   the Galilean symmetry. This new solution
is obtained by the use of Null Melvin Twist (NMT) or the Discreet Light Cone  
Quantisation (DLCQ). One of the difference between these two techniques 
 is that in the later case the 
light cone direction need to be compact, which means the momentum can not
be any more continuous, whereas in the earlier case we 
do not require this to be the case otherwise we won't be able boost along
the this direction and the momentum is continuous. 
Because of this compact nature of the light cone 
direction,  special care need 
to be required for the zero modes \cite{my}, for other important
references see \cite{abs} and \cite{mmt} and references therein.

In this paper we shall employ the earlier approach i.e. the null Melvin twist
to generate new gravitational solution from the known solution. The effect of 
NMT is to generate a solution which is of the plane wave type \cite{ghhlr} and 
some aspects of the dual field theory  has been studied in \cite{ag}.

In this context,  the solution generating technique \cite{ghhlr} 
 has been applied to 
the super conformal  solution of D3 brane  and new solutions has been obtained
and has been interpreted as the solution showing non-relativistic 
conformal symmetry \cite{hrr} and \cite{abm}. This solution generating
technique is becoming interesting  due to the appearance of Galilean 
(conformal) invariance and is believed to be useful in the understanding of the
strongly coupled behaviour of some condensed matter systems. Recently, the
explicit form of a solution has been constructed which preserves this
symmetry \cite{son} and \cite{bm} and  this solution has been embedded in string
theory in \cite{mmt}, \cite{hrr} and  \cite{abm} using the above mentioned
techniques.

There has been several other important, interesting and independent 
studies by other groups \cite{ch},\cite{msw},\cite{ns},\cite{wg}, \cite{bf} 
and references therein.

Recently, there appeared another solution \cite{klm} in 3+1 dimension, 
which has an intrinsic exponent z in a coordinate independent way
i.e. the Ricci scalar, square of Ricci tensor $R_{MN}R^{MN}$  and the
square of Riemann tensor $R_{MNKL}R^{MNKL}$ depends on this exponent z,
whereas the metric written in \cite{son},\cite{bm}, \cite{abm},\cite{hrr}
do depends on this exponent but not the coordinate invariant quantities.
Probably, we can say that both of these kind of solutions do not fall in the
same class. Another way to see these solutions are not of the same type is 
that one type of solutions
are of plane wave type whereas \cite{klm} is not. It is interesting to note
that for z=1 both of these kind of solutions satisfy the relation $R_{MN}\sim
g_{MN}$, whereas as soon as we go away from z=1 then none of these solutions
preserves the relation $R_{MN}\sim g_{MN}$. Which  we  may interpret as that 
both of these class of solutions are non-maximally symmetric space times for 
z away from 1. We can generalize the solution of \cite{klm} by introducing
another exponent and study the correlation function \cite{ssp}. 
   
The NMT has already been applied to  systems for odd value of p of Dp brane
which are charged under electric field  in \cite{nps} and some other
systems using different techniques  in \cite{bfhp}-\cite{hrr1} and some other 
aspects of the plane wave solutions has been studied in \cite{st}-\cite{zs}
and references therein.

Here we are interested to study a daughter field theory via gauge gravity 
duality whose parent field theory exhibits
some interesting properties, namely, shows confinement and upon inclusion of
flavors it shows chiral symmetry breaking and at finite temperature the 
system restores the chiral symmetry.
    
The study of strongly coupled field theory via AdS/CFT correspondence 
has attracted a lot interest and attention. In this context, the obvious 
thing to try and understand is
the study of the strongly coupled behavior of QCD, but to understand 
 QCD  via the above mentioned 
correspondence means we  
are studying it in the large but fixed 't Hooft coupling and   at large
rank of the gauge group. More importantly, finding a suitable
 geometric description of QCD has been very non-trivial.  
Moreover, QCD contains flavor degrees of freedom, which transforms fundamentally
under the color gauge group, 
 it means from the holographic dual bulk theory point of view, we need to 
 include flavors and there exists an approach to incorporate  and 
 fix the arbitrariness of the number of flavors. 

One such approach is called the probe-brane approximation, where one 
considers the limit in which the number branes that produces the 
gluonic degrees of freedom, called color branes, are large in 
comparison to the number of flavors \cite{kk}. 
It means the flavor branes do not 
back react on the color branes and hence the presence of flavor branes 
are not going to change the background geometry  generated by the color brane.
In order to understand the physics of these flavors more accurately,  one need
to  consider the full 
back reaction of the flavor branes onto the color brane.

The studies  to understand several properties of 
QCD till now is  either by the top-down \cite{ss} or the bottom-up approaches
\cite{ekss}, \cite{gk} and \cite{gkn}. 
In the former case
one starts with a theory which has got either 
the right color gauge group, matter
content and couplings or it shares some features with QCD. 
One such example is the Sakai-Sugimoto model, which displays  
confinement and chiral symmetry breaking, and   
this particular approach has got its own problem, like one has to deal
here via the probe brane approximation,  an unwanted SO(5) symmetry group, 
and the most trouble some 
unwanted complication is the presence of KK modes\footnote{Some studies have
been made in \cite{ssp1} using the solution generating technique \cite{lm}, 
but more interesting studies need to be made.}.  
However, it is not the case
with the bottom-up approach, where one starts with a theory with a minimal 
matter content and the  gauge group and assumes that it has the 
correct coupling and then  proceeds to calculate several interesting 
quantities like chiral symmetry breaking and current-current correlators 
\cite{ekss} using AdS/CFT correspondence. 

From the days of AdS/QCD,  several interesting studies have been made to 
understand the holographic QCD better. In this context, the IR properties like
chiral symmetry breaking, its restoration, confinement and the transition 
between the confining and de-confining phases    
have been understood  \cite{ahjk}-\cite{gp} from the top-down approach,
than its UV properties like the asymptotic freedom \cite{gk}, which is 
studied using the bottom-up approach. 
Since these are the most important properties of QCD, means we should have 
them when we construct any model holographically, at least it should  fall 
in the same class of QCD by  showing  these properties. Until, now it has
not been possible to embed all these properties into  a  single model
from top-down approach. The vev of the bi-fundamental field associated to the
chiral symmetry breaking is studied in \cite{ak}, \cite{hhly} and \cite{nms}
 using open Wilson line and tachyon condensation
 in \cite{ckp}, \cite{bss} and \cite{dn}.  

In this paper, we would like to understand better the Sakai-Sugimoto model.
In particular, applying the Null Melvin Twist (NMT) to the near horizon 
limit of D4 brane solution, generates a solution that asymptotes to a plane 
wave solutions and would study different aspects of Sakai-Sugimoto model
like: how does this deformation to the original solution changes confining
de-confining transition and also the chiral symmetry breaking and restoration
curve? 

The results of this study can be summarized as:\\

(1)The application of NMT to the near extremal solution breaks
the rotational symmetry SO(5)  transverse to the D4 brane. However, it does not 
break the SO(5) symmetry of the zero temperature solution.\\

(2)Even though there is no rotation in the NMT resulted solution but there
appears the angular dependence in the metric component parallel to the brane
direction. The consequence of this is that the scalar field moving 
in this geometry
is not anymore just a function of the radial coordinate at zero frequency and 
zero momentum, which makes the 
 computation of the correlation functions very complicated. \\

(3) As a result the coefficient of shear viscosity is not any more given by the
formula eq(3.6) of hep-th/0309213.\\ 

(4) The NMT resulted solution to the D4 brane has got the same four-form 
field strength for both zero and non-zero temperature solution. The dilaton 
for the non-zero temperature  is different whereas at zero temperature 
it remains same.\\

(5)The periodicity associated to Euclidean time circle for the black hole
solution and the $\tau$ circle for the zero temperature solution are same as
before the application of NMT.\\

(6)The curve that describes the chiral symmetry breaking and restoration
\cite{asy},\cite{gp} remains unchanged.\\

(7)The curve for confining and de-confining transition \cite{asy},\cite{gp}
 also remain unchanged.

\section{The D4 brane solution}

Let us consider a
 model that talks about generating  holographically a gravity solution which 
may fall in the same class as that of QCD as far as only  confinement and 
chiral symmetry breaking is concerned. The model is based on the D4 brane 
solution with no supersymmetry, in string frame it reads 
\bea
ds^2_s&=&(\f{r}{R})^{3/2}[-dt^2+\sum^3_1 dx^i dx_i+f(r) d\tau^2]+
(\f{R}{r})^{3/2}[dr^2+r^d\Omega^2_4],\nn 
\Phi&=&Log[g_s(\f{r}{R})^{3/4} ],~~~F_4=\f{2\pi N_c}{v_4}\epsilon_4,~~f(r)=
1-(\f{r_0}{r})^3
\eea 
where $v_4=\f{8\pi^2}{3}$, $\epsilon_4$ is the volume form of the unit 
$S^4$ and $R^3=\pi g_s N_c l^3_s$. Note that only this form of normalization
 of the $F_4$ flux satisfy the necessary
flux quantization condition in units of $2\pi l_s=1$, which is 
\be
\int_{S^4} \star_{10}F_6=\int_{S^4} F_4=2\pi N_c
\ee 

and if we do not work in any such units then $v_4=\f{2\pi}{3l^3_s}$ and the 
normalization condition is
\be
\int_{S^4} \star_{10}F_6=\int_{S^4} F_4=(4\pi^2\alpha')^{\f{3}{2}}N_c
\ee 

This background is reliable for a certain range of energy scale \cite{imsy},
 for which the effective coupling in field theory 
should be very large or in other words the 
Ricci scalar should be very small in units of string length, $l_s$. To 
suppress the effects of string loop corrections, we need to take small 
values to dilaton, $e^{\Phi} \ll 1$.

According to gauge gravity correspondence, the five dimensional coupling
is related to a dimensionless effective coupling $g_{eff}$ \cite{imsy}
\be g^2_{eff}=g^2_5 N_c U ,~~~{\rm with} ~~g^2_5=(2\pi)^2 g_s l_s,
\ee
where $U=\f{r}{\alpha'}$ and the four dimensional 't Hooft coupling is 
related to the five dimensional coupling
\be
g^2_{YM}N_c=\f{g^2_5 N_c}{2\pi R_{\tau}},
\ee
where $2\pi R_{\tau}$ is the periodicity of the $\tau$ circle and is written
in eq(\ref{periodicities}). It follows that the four dimensional 't Hooft
coupling is $g^2_{YM}N_c= \f{3}{l^2_s} R^{3/2}\sqrt{r_0}$. From the
D4 solution, the Ricci scalar gets a maximum value at $r=(\f{7}{5})^{1/3}r_0$
and is 
\be 
R_s=-\f{135}{14} \f{(5/7)^{1/6}}{R^{3/2}\sqrt{r_0}}
\ee 
Demanding that $\alpha' |R_s| \ll 1$ gives 
\be
\f{R^{3/2}\sqrt{r_0}}{\alpha'}\gg 1.
\ee
which means the four dimensional 't Hooft coupling is large i.e. 
$g^2_{YM}N_c \gg 1$. So, roughly, its the energy that stay close to IR
makes the gravity description reliable.  

 From the dilaton constraint we get 
\be
U\ll \f{4\pi^{7/3} N^{4/3}}{g^2_{YM}N_c 2\pi R_{\tau}}=
\f{3 \pi^{1/3}N^{1/3}}{g_s l^3_s}
\ee
\section{Null Melvin Twist}
We would like to do Null Melvin Twist (NMT) to backgrounds of the form 
\be\label{d4_sol} 
ds^2=A(r) dt^2+B(r) dy^2+\delta_{ij}(r)dx^idx^j+ C(r) dr^2+ D(r) d\Omega^2_4
\ee
There could be other fields present that are coming either from the NS-NS or
RR sector or both. As a particular example, we shall stick to the D4 brane
geometry, for which there are not any other fields coming from the NS-NS sector 
apart from dilaton.
The dilaton 
\be \Phi(r)=\Phi_0(r)\ee
and the 4-form flux is taken as
\be F_4=f_0 \epsilon_4,\ee
where $f_0$ is some normalization constant and $\epsilon_4$ is the volume 
form of the 4-sphere.
 
Later we shall decompose the metric of $d\Omega^2$ explicitly 
when we do the twist along one of its isometry direction.

The proposal of doing Null Melvin twist consists of six steps \cite{ghhlr}.

1. Boost along a translationally invariant direction, y, by an amount $\gamma$.

2. T-dualize along this direction, y.

3. Twist a one form i.e. do  rotations in different planes

4. T-dualize again on y.

5. Boost by -$\gamma$ along the same direction, y

6. Take a specific limit in which $\alpha\rightarrow 0$, 
$\gamma\rightarrow\infty$ and keeping $\beta=\f{1}{2}\alpha e^{\gamma}$=
fixed  

By going through these 6 steps, we generate a new background and in the 
$\beta \rightarrow 0$, we get back our original solution. 
 
Notation: c = cosh~$\gamma$,  s = sinh~$\gamma$

Let us start to do the NMT for the geometry written above\footnote{Sometimes 
we write A(r) and B(r) etc, as simply A and B etc, just to avoid cluttering of
brackets. }.

Step 1: 
\be dt\rightarrow c dt- s dy,~~~~~dy\rightarrow -s dt+c dy\ee 
\bea
ds^2&=&(A c^2+B s^2) dt^2 +(A s^2+B c^2) dy^2-2 cs (A+B) dt dy+\nn 
&&\delta_{ij}(r)dx^idx^j+ C(r) dr^2+ D(r) d\Omega^2_4,\nn 
&& \Phi(r)=\Phi_0(r),~~  F_4=f_0 \epsilon_4.
\eea

Step 2:
\bea
ds^2&=&\f{AB}{(A s^2+B c^2)} dt^2+\f{dy^2}{(A s^2+B c^2)}+\delta_{ij}(r)dx^idx^j+ C(r) dr^2+ D(r) d\Omega^2_4,\nn && \Phi(r)=\Phi_0(r)-\f{1}{2}Log[(A s^2+B c^2)],B_{ty}=-\f{cs (A+B)}{(A s^2+B c^2)},\nn &&  F_5=f_0 dy\w\epsilon_4
\eea

In order to perform step 3, let us write down the metric of unit 4-sphere as
\be d\Omega^2_4=dx^2_1+\cdots+dx^2_5,\ee with the restriction 
$x^2_1+\cdots+x^2_5=1$. The twisting is done following \cite{ghhlr}

\bea x_1+i x_2 &\rightarrow& e^{i\alpha y} (x_1+i x_2)\nn
  x_3+i x_4 &\rightarrow& e^{i\alpha y} (x_3+i x_4),\eea

where we have done equal rotation in both the planes. This gives the 
resulting unit 4-sphere metric 
\be d\Omega^2_4\rightarrow d\Omega^2_4+\alpha \sigma dy+\alpha^2(1-x^2_5) dy^2,
\ee  
where \be \f{\sigma}{2}=x_1 dx_2-x_2 dx_1+x_3dx_4-x_4dx_3.\ee 

A specific realization of unit $S^4$ could be 
\bea &&x_1= sin\theta~sin\varphi~sin\psi,~~x_2= sin\theta~sin\varphi~cos\psi\nn&& x_3= sin\theta~cos\varphi~sin\chi,~~x_4= sin\theta~cos\varphi~cos\chi\nn
&& x_5=cos\theta,
\eea 
with the ranges for the angles are $0\leq \theta \leq \pi,~~ 0\leq \varphi\leq 
\f{\pi}{2},~~0\leq \psi,~\chi\leq 2\pi$.

It means the resulting unit 4-sphere metric with this realization becomes
\be d\Omega^2_4\rightarrow d\Omega^2_4+\alpha \sigma dy+\alpha^2 sin^2\theta dy^2 \ee

The one form 
\be \f{\sigma}{2}=-sin^2\theta sin^2\varphi d\psi -
sin^2\theta cos^2\varphi d\chi\ee

and the \be d\Omega^2_4=d\theta^2+sin^2\theta d\Omega^2_3,\ee
where \be d\Omega^2_3=d\varphi^2+sin^2\varphi d\psi^2+cos^2\varphi d\chi^2\ee
Due to this twisting the volume form $\epsilon_4$ will be different. This can
be seen as follows.

The metric of the unit 4-sphere can be re-written as 
\bea ds^2_4&=&\f{1}{1-\sum^4_1 x^2_i} \Bigg((1-\sum^4_1 x^2_i+x^2_1)dx^2_1+(1-\sum^4_1 x^2_i+x^2_2)dx^2_2+\nn&&(1-\sum^4_1 x^2_i+x^2_3)dx^2_3+(1-\sum^4_1 x^2_i+x^2_4)dx^2_4+2 x_1 x_2 dx_1 dx_2+2 x_1 x_3 dx_1 dx_3+\nn&&2 x_1 x_4 dx_1 dx_4+2 x_2 x_3 dx_2 dx_3+2 x_2 x_4 dx_2 dx_4+2 x_3 x_4 dx_3 dx_4\Bigg)\eea

The volume form of this metric is 
\be\label{volume_form} \epsilon_4= \f{1}{\sqrt{1-\sum^4_1 x^2_i}}dx_1\w dx_2\w dx_3\w dx_4,\ee
expressing this in terms of the angular coordinates give\be 
\epsilon_4=sin^3\theta sin\varphi cos\varphi d\theta\w d\varphi\w d\psi\w d\chi
\ee  

Now using the following fact 
\bea 
dx_1+i dx_2 &\rightarrow& e^{i\alpha y} [(dx_1+idx_2)+i \alpha dy (x_1+i x_2)]\nn dx_3+i dx_4 &\rightarrow& e^{i\alpha y} [(dx_3+idx_4)+i \alpha dy 
(x_3+i x_4)]\eea
we get 
\bea &&dx_1\w dx_2\w dx_3\w dx_4\rightarrow dx_1\w dx_2\w dx_3\w dx_4-
\nn&&\alpha dy\w [(x_3 dx_3+x_4 dx_4)\w dx_1\w dx_2+(x_1 dx_1+x_2 dx_2)\w dx_3
\w dx_4]\nn
\eea

This means the volume form changes under the twisting to
\be  \epsilon_4 \rightarrow \epsilon_4- \alpha sin^3\theta sin\varphi 
cos\varphi ~dy\w ~d\theta\w ~d\varphi\w~ d(\chi-\psi)\ee  
It means under this twisting the five form flux generated after step 2, 
do not gets changed.

Step 3:
After the twist the solution looks as
\bea
ds^2&=&\f{AB}{(A s^2+B c^2)} dt^2+\f{dy^2}{(A s^2+B c^2)}+\delta_{ij}(r)dx^idx^j+ C dr^2+ D d\Omega^2_4+\nn &&D\alpha \sigma dy+D\alpha^2 sin^2\theta dy^2 , \Phi(r)=\Phi_0(r)-\f{1}{2}Log[(A s^2+B c^2)],\nn &&B_{ty}=-\f{cs (A+B)}{(A s^2+B c^2)},  F_5=f_0 dy\w \epsilon_4,\nn&&\Phi(r)=\Phi_0(r)-\f{1}{2}Log[(A s^2+B c^2)]
\eea

Step 4:

T-dualizing the solution along y direction that resulted after step 3 gives
\bea 
ds^2&=&X dt^2+W dy^2+2 Z dt dy+ \delta_{ij}(r)dx^idx^j+ C dr^2+D d\theta^2+ 
D sin^2\theta d\varphi^2+\nn && L d\psi^2+M d\chi^2+2 N d\chi d\psi,\\ 
\Phi &=&\Phi_0-\f{1}{2}Log[K], F_4=f_0\epsilon_4\nn
B&=&-\f{D\alpha sin^2\theta}{K}[sin^2\varphi(A s^2+B c^2)d\psi\w dy+
cos^2\varphi(A s^2+B c^2)d\chi\w dy+\nn&&sin^2\varphi c s(A +B)dt\w d\psi+
cos^2\varphi cs(A+B)dt\w d\chi],
\eea
where 
\bea
K& = &1+D \alpha^2 sin^2\theta(As^2+Bc^2),\nn
X& = &\f{AB K+c^2 s^2 (A+B)^2}{K(As^2+Bc^2)},\nn W &=&\f{As^2+Bc^2}{K},~~~
Z=-\f{cs(A+B)}{K},\nn
L& = &\f{Dsin^2\theta sin^2\varphi[1+D\alpha^2sin^2\theta cos^2\varphi(As^2+Bc^2)]}{K},\nn M&=&\f{Dsin^2\theta cos^2\varphi[1+D\alpha^2sin^2\theta sin^2\varphi(As^2+Bc^2)]}{K},\nn  N & = &-\f{D^2 \alpha^2 sin^4\theta sin^2\varphi cos^2\varphi}{K}(A s^2+B c^2)
\eea

Step 5:
Boosting the solution after step 4 by an amount $-\gamma$ gives 
\bea 
ds^2&=&(c^2 X+ s^2 W +2 cs Z) dt^2+(s^2 X+ c^2 W +2 cs Z) dy^2+\nn&&
2 dtdy[cs(X+W)+(c^2+s^2)Z]+\delta_{ij}(r)dx^idx^j+ C dr^2+D d\theta^2+\nn && 
D sin^2\theta d\varphi^2+ L d\psi^2+M d\chi^2+2 N d\chi d\psi,\nn 
\Phi &=&\Phi_0-\f{1}{2}Log[K],~~~ F_4=f_0 \epsilon_4\nn
B&=&\f{D\alpha sin^2\theta}{K}[-sin^2\varphi A s dt\w d\psi+B c sin^2\varphi dy\w d\psi-cos^2\varphi A s dt\w d\chi\nn&&+cos^2\varphi c B  dy\w d\chi]
\eea

Step 6:

The final step is to take a proper limit for which  $\alpha \rightarrow 0$ and
$\gamma\rightarrow\infty$ with $\beta=\f{\alpha}{2}e^{\gamma}$ fixed, gives 

\bea\label{gen_metric} ds^2&=&\f{A}{K}(1+\beta^2 DB sin^2\theta)dt^2+\f{B}{K}(1+\beta^2 DA sin^2\theta)dy^2+2 \f{ABD\beta^2 sin^2\theta}{K}dtdy\nn&+&\delta_{ij}(r) dx^i dx^j+C dr^2+D d\theta^2+D sin^2\theta d\varphi^2\nn&+&\f{D}{K} sin^2\theta sin^2\varphi(1+\beta^2 D sin^2\theta cos^2\varphi (A+B))d\psi^2\nn&+&\f{D}{K} sin^2\theta cos^2\varphi (1+\beta^2 D sin^2\theta sin^2\varphi(A+B))d\chi^2\nn&-&2 \f{\beta^2 D^2 sin^4\theta sin^2\varphi cos^2\varphi(A+B)}{K}d\chi d\psi,
\nn
B_2&=&\f{D\beta sin^2\theta}{K}[-sin^2\varphi A  dt\w d\psi+B  sin^2\varphi dy\w d\psi-cos^2\varphi A  dt\w d\chi\nn&&+cos^2\varphi  B  dy\w d\chi],\nn
\Phi &=&\Phi_0-\f{1}{2}Log[K],~~~ F_4=f_0 \epsilon_4,~~K=1+\beta^2 D sin^2\theta(A+B)\eea

The  simplified form of the NMT resulted D4 brane geometry can be written as
\bea ds^2&=& \f{A}{K}dt^2+\f{B}{K}dy^2+\f{ABD\beta^2 sin^2\theta}{K}(dt+dy)^2+\delta_{ij}(r) dx^i dx^j+C dr^2\nn&+&D d\theta^2+D sin^2\theta d\varphi^2+
\f{D}{K} sin^2\theta sin^2\varphi d\psi^2+\f{D}{K} sin^2\theta cos^2\varphi d\chi^2\nn&+&\f{\beta^2}{K}D^2sin^4\theta sin^2\varphi cos^2\varphi(A+B) (d\psi-d\chi)^2 \eea

Upon doing the following change of coordinates,  
$t=\f{1}{2}({\tilde t}-\xi)$,~~ $y=\f{1}{2}({\tilde t}+\xi)$, we can rewrite
the metric in a form that shows explicitly the pp-wave structure.

\bea ds^2&=& (\f{A+B}{4K}+\beta^2 \f{ADB}{K} sin^2~\theta)d{\tilde t}^2+
\f{A+B}{4K}d\xi^2
+2\f{B-A}{4K} d{\tilde t} d\xi+\delta_{ij}(r) dx^i dx^j\nn&+&C dr^2+D d\theta^2+D sin^2\theta d\varphi^2+
\f{D}{K} sin^2\theta sin^2\varphi d\psi^2+\f{D}{K} sin^2\theta cos^2\varphi d\chi^2\nn&+&\f{\beta^2}{K}D^2sin^4\theta sin^2\varphi cos^2\varphi(A+B) (d\psi-d\chi)^2,\nn 
B_2&=& \f{D\beta sin^2\theta }{2K}[-A(d{\tilde t}-d\xi)+B(d{\tilde t}+d\xi)]\w [sin^2\varphi d\psi+cos^2\varphi d\chi] \eea
 
For the example at hand the functions are
\bea\label{bh_functions} 
A&=&-f(r) (\f{r}{R})^{3/2},~~B(r)=(\f{r}{R})^{3/2},~~\delta_{ij}(r)=\delta_{ij}(\f{r}{R})^{3/2},~~C(r)=(\f{R}{r})^{3/2} \f{1}{f(r)},\nn 
D(r)&=&R^{3/2}\sqrt{r},~~
f(r)=1-(\f{r_T}{r})^3,~~\Phi_0=Log[g_s (\f{r}{R})^{3/4}],~~
f_0=\f{2\pi N_c}{v_4}, 
\eea
  which describes a black hole and $v_4$ is the unit volume of $S^4$, whose 
value is $v_4=\f{8\pi^2}{3}$.   Re-writing the solution in a better looking 
 way 
\bea\label{solI} 
ds^2&=&\f{1}{K}(\f{r}{R})^{3/2}\bigg[-f(1+r^2 \beta^2 sin^2\theta)dt^2+
(1-fr^2 \beta^2 sin^2\theta)dy^2-2f r^2 \beta^2 sin^2\theta dt dy\nn&&+K(dx^2+dz^2+d\tau^2)\bigg]+(\f{R}{r})^{3/2}\f{dr^2}{f}+\f{\beta^2}{K}\f{R^{3/2} r^3_0}{\sqrt{r}} sin^4\theta sin^2\varphi cos^2\varphi (d\psi-d\chi)^2\nn&&+\f{R^{3/2}\sqrt{r}}{K}\bigg[K(d\theta^2+sin^2\theta d\varphi^2)+sin^2\theta(sin^2\varphi d\psi^2+cos^2\varphi d\chi^2)\bigg]\nn
B_2&=& \f{\beta r^2}{K}sin^2\theta (fdt+dy)\w[sin^2\varphi d\psi+
cos^2\varphi d\chi],\nn
\Phi &=&\Phi_0-\f{1}{2}Log[K],~~~ F_4=f_0 \epsilon_4,~~K=1+\f{r^3_T}{r}\beta^2  sin^2\theta
\eea

In this case it follows that the original SO(5) symmetry transverse to the 
D4 brane is broken to 
$SO(3)\times U(1)^2$. The SO(3) is spanned by the $\theta, ~\varphi$ coordinates
and the two U(1)'s follow from $\chi$ and $\psi$ coordinates.

Let us do a double Wick rotation to the solution that is presented in 
eq(\ref{d4_sol}), then the resulting solution shows a  very interesting 
property that is  confinement. 

If the functions are
\bea\label{ext_functions}
A&=&- (\f{r}{R})^{3/2},~~B(r)=(\f{r}{R})^{3/2},~~\delta_{11}(r)=\delta_{22}(r)=(\f{r}{R})^{3/2},\delta_{33}(r)=f(r)(\f{r}{R})^{3/2},
\nn C(r)&=&(\f{R}{r})^{3/2} \f{1}{f(r)},~~D(r)=R^{3/2}\sqrt{r},~~
f(r)=1-(\f{r_0}{r})^3,~~\Phi_0=Log[g_s (\f{r}{R})^{3/4}],\nn
f_0&=&\f{2\pi N_c}{v_4},
\eea 
then it describes a system at zero temperature which shows confinement for 
$\beta=0$. Note that for non-zero $\beta=\pm \f{cosec~\theta_0}{r_0 }$, 
the surface
$(r,~\theta)=(r_0,~\theta_0)$ shows that the $g_{yy}$ goes to zero.  
This is just an artifact of the coordinate system we used.

Note in this case $A+B=0$, which means the function $K=1$ and the solution can 
be written in a better way as
\bea\label{solII}
ds^2&=&(\f{r}{R})^{3/2}\Bigg[-(1+r^2 \beta^2 sin^2\theta)dt^2+(1-r^2 \beta^2 sin^2\theta)dy^2-2 r^2 \beta^2 sin^2\theta dt dy+\nn&&(dx^2+dz^2+fd\tau^2)\Bigg]+(\f{R}{r})^{3/2} \f{dr^2}{f(r)}+R^{3/2}\sqrt{r}d\Omega^2_4,\nn
B_2&=&r^2\beta sin^2\theta (dt+dy)\w[sin^2\varphi d\psi+
cos^2\varphi d\chi],\nn \Phi&=&\Phi_0,~~F_4=f_0 \epsilon_4.
\eea
    
In this case the symmetry transverse to the D4 brane is not broken by the 
NMT twist that is the SO(5). 
It is very easy to conclude that eq(\ref{solII}) cannot be generated from 
eq(\ref{solI}) by a double Wick rotation and it implies that  NMT do not
commute with  Wick rotation.

Let us do some change of coordinates $t=\f{1}{2}({\tilde t}-\xi)$,~~ 
$y=\f{1}{2}({\tilde t}+\xi)$ and $\rho=2 \f{R^{3/2}}{\sqrt{r}}$, 
using these we can rewrite the solution 
eq(\ref{solII})
\bea
ds^2&=&ds^2_5+(\f{r}{R})^{3/2}fd\tau^2+R^{3/2}\sqrt{r}d\Omega^2_4,~~\Phi=\Phi_0,~~F_4=f_0 \epsilon_4,\nn
B_2&=&4\sqrt{2}\beta R^3\f{sin^2\theta}{\rho^2}d{\tilde t}\w [sin^2\varphi 
d\psi+cos^2\varphi d\chi],
\eea
with the
\be
ds^2_5={\tilde R}^2
\bigg[\f{-2 \tilde{\beta}^2 {\cal S}(\theta) d{\tilde t}^2}{\rho^{2z+\nu}}+
\f{2d{\tilde t}d\xi+dx_idx_i+d\rho^2}{\rho^{2+\nu}}\bigg],
\ee
where
\be 
z=3,~~~\nu=1, ~~~{\cal S}(\theta)=sin^2\theta,~~~{\tilde R}^2=8 R^3,~~{\tilde \beta}^2=2^4 R^{6}\beta^2.
\ee

Now we have got two ``exponents'' $z$ and $\nu$ along with a function 
${\cal S}(\theta)$.

As a comparison with an extremal D3 brane \cite{bm} 
\be
z=2,~~\nu=0, ~~{\cal S}(\theta)=1.
\ee

Now we may say  that the field theory living in 3+1 dimension
 possessing the Galilean 
symmetry should have a five dimensional geometry  
\be
ds^2_5={\tilde R}^2
\bigg[\f{-2 \tilde{\beta}^2 {\cal S}(\theta) d{\tilde t}^2}{\rho^{2z+\nu}}+
\f{2d{\tilde t}d\xi+dx_idx_i+d\rho^2}{\rho^{2+\nu}}\bigg]
\ee

The exponents $\nu$ and $z$ are both  ``dynamical''  in the sense that 
as it depends on the 
dimensionality of the brane. Hopefully, it will appear in the realistic 
sense i.e. in the result of the correlation function 
\be 
z=\f{7-p}{5-p},~~~\nu=\f{p-3}{5-p}.
\ee
This is true for any Dp brane with an exception to D5 brane.

Summarizing, the above analysis it follows that 
 the more feasible looking interpretation would be that
 the part of the 5d geometry, on the boundary  of whose the field theory
lives, in general can have two exponents $z$ and $\nu$ which depends
 on p  of Dp brane and it admits non-relativistic non-conformal
 symmetry and for p=3 it shows as usual the non-relativistic
  conformal symmetry.

The above classification is independent of whether  the  ten dimensional  
solution  have a piece that  admit a Sasaki-Einsteinian structure or not.  

One interesting point to note that the solution eq(\ref{solI}) has two 
limits $\beta\rightarrow 0$ and $r_T\rightarrow 0$. In the former limit 
it goes over to the original solution that we started out with i.e. 
the non-extremal black D4 brane solution.  In the
latter limit we get a solution which is the result of applying 
Null Melvin twist to the extremal D4 brane solution. 
 
For completeness, let us summarize all the solutions, which may be 
of importance in the study of the AdS/QCD from top-down approach
\footnote  {As this is the only system 
available where one can talk of many interesting aspects like chiral
symmetry breaking and confinement}

Since the $r_T\rightarrow 0$ limit gives us an extremal solution means
this is not of much important to us in the present study, however could
be very interesting to understand the mesonic and glue ball spectrum in the
light of study of AdS/QCD.
We shall take only the $\beta\rightarrow 0$ limit of solutions 
eq(\ref{solI}) and eq(\ref{solII}) for our further study. 

Note that the solution eq(\ref{solI}) is not a rotating solution as the 
coordinate y is non-compact. It is also very surprising to see that 
the metric components  depends on the angular coordinate $\theta$, even  
without rotation.

Is this generic to Dp brane solution with an even dimensional sphere, in this
case $S^4$, transverse to the brane directions ? The answer is yes.

Just to see that let us write down the metric of a D dimensional sphere 
\be d\Omega^2_D=dx^2_1+dx^2_2+\cdots\cdots+dx^2_D
\ee
with a restriction $x^2_1+x^2_2+\cdots\cdots+x^2_D=1$. Here we can have
D=2d or 2d+1, even or odd dimensional sphere. 
For a generic D dimensional case we can 
have d number of independent planes so as to make rotations in each plane. 
for Simplicity, we shall consider the same amount of rotation in each plane.

Let us do a rotation of (say) $x_j$ and $x_{j+1}$ plane as\be
x_j+x_{j+1}\rightarrow e^{i\alpha y}[x_j+i x_{j+1}]\ee where $y$ is the 
isometry direction along which we have done the boosting.  This rotation means
\be dx^2_j+dx^2_{j+1}\rightarrow  dx^2_j+dx^2_{j+1}+\alpha^2(x^2_j+x^2_{j+1}) 
dy^2 +2 \alpha dy (x_j dx_{j+1}-x_{j+1}dx_j)\ee

Now it easily follows that the metric of a D=2d dimensional sphere, after the
rotation \be d\Omega^2_{2d}=d\Omega^2_{2d}+\alpha^2 dy^2+2\alpha(x_1dx_2-x_2dx_1+\cdots\cdots+x_{2d-1}dx_{2d}-x_{2d}dx_{2d-1})dy\ee

whereas for an odd dimensional sphere, $D=2d+1$, the metric after rotation 
\be 
d\Omega^2_{2d+1}=d\Omega^2_{2d+1}+\alpha^2(1-x^2_{2d+1}) dy^2+2\alpha(x_1dx_2-x_2dx_1+\cdots+x_{2d-1}dx_{2d}-x_{2d}dx_{2d-1})dy
\ee

The coefficient of $g_{yy}$ of the sphere part of the metric is going to 
appear all over the places once we do  T-duality along this direction
following the  rules of Null Melvin Twist. Hence the appearance of the 
angular variable on the metric component is a must.

The temperature of NMT resulted  black hole solution eq(\ref{solI}) is same
 as the one before applying NMT to the corresponding black hole solution. 
Similarly,  the 
periodicity of the extremal solution eq(\ref{solII}) is same as the solution
before applying Null Melvin Twist. Which means the  temperature of the 
black hole solution and the
periodicity of the compact circle $\tau$ of the extremal solution 
are independent of the parameter $\beta.$

In order to see these, first we have to compute the surface gravity 
\be
\kappa^2=-\f{1}{2}(\nabla^a \varepsilon^b)(\nabla_a \varepsilon_b)
\ee
where $\varepsilon^b$ is a null Killing vector defined on the horizon.
The  temperature of the black hole is defined as 
$T_{H}=\f{\kappa}{2\pi}$. 

Similarly to compute the periodicity of the circle $\tau=x^4$, we define 
an analogous quantity ``surface gravity" but this time without the minus
sign 
\be
\kappa^2_{x^4}=\f{1}{2}(\nabla^a \epsilon^b)(\nabla_a \epsilon_b)
\ee
where $\epsilon^b$ is a null Killing vector defined on the surface
for which $g_{x^4x^4}$ vanishes.
The inverse periodicity  of this circle is defined as 
$\f{1}{2\pi R_{\tau}}=\f{\kappa_{x^4}}{2\pi}$. 

Taking $\varepsilon^b=\left(\f{\p}{\p t}\right)^b$ and  
$\epsilon^b=\left(\f{\p}{\p x^4}\right)^b$, we get 
\be 
\label{periodicities}
T_H=\f{3}{4\pi} \sqrt{\f{r_T}{R^3}},~~~2\pi R_{\tau}=\f{4\pi}{3} 
\sqrt{\f{R^3}{r_0}}
\ee

Another interesting thing is that the entropy of the black hole solution 
eq(\ref{solI}) do not depends on the twist parameter $\beta.$ This is 
in agreement with \cite{ghhlr}. Note that the solution eq(\ref{solI}) is
written in string frame.

For completeness the solution of the black hole in Einstein frame is
\bea
\label{solI_e} 
ds^2_E&=&\f{1}{K^{3/4}}(\f{r}{R})^{9/8}\bigg[-f(1+r^2 \beta^2 sin^2\theta)dt^2+
(1-fr^2 \beta^2 sin^2\theta)dy^2\nn&-&2f r^2 \beta^2 sin^2\theta dt dy+K(dx^2+dz^2+d\tau^2)\bigg]+(\f{R}{r})^{15/8}K^{1/4}\f{dr^2}{f}\nn&+&\f{\beta^2}{K^{3/4}}\f{R^{15/8} r^3_0}{r^{7/8}} sin^4\theta sin^2\varphi cos^2\varphi (d\psi-d\chi)^2\nn&&+\f{R^{15/8} r^{1/8}}{K^{3/4}}\bigg[K(d\theta^2+sin^2\theta d\varphi^2)+sin^2\theta(sin^2\varphi d\psi^2+cos^2\varphi d\chi^2)\bigg]\nn
B_2&=& \f{\beta r^2}{K}sin^2\theta (fdt+dy)\w[sin^2\varphi d\psi+
cos^2\varphi d\chi],\nn
\Phi &=&\Phi_0-\f{1}{2}Log[K],~~~ F_4=f_0 \epsilon_4,~~K=1+\f{r^3_T}{r}\beta^2  sin^2\theta
\eea

and the extremal solution is

\bea\label{solII_e}
ds^2_E&=&(\f{r}{R})^{9/8}\Bigg[-(1+r^2 \beta^2 sin^2\theta)dt^2+(1-r^2 \beta^2 sin^2\theta)dy^2-2 r^2 \beta^2 sin^2\theta dt dy+\nn&&(dx^2+dz^2+fd\tau^2)\Bigg]+(\f{R}{r})^{15/8} \f{dr^2}{f(r)}+R^{15/8} r^{1/8}d\Omega^2_4,\nn
B_2&=&r^2\beta sin^2\theta (dt+dy)\w[sin^2\varphi d\psi+
cos^2\varphi d\chi],\nn \Phi&=&\Phi_0=Log[g_s (\f{r}{R})^{3/4}],~~
F_4=f_0 \epsilon_4.
\eea

\section{Chiral symmetry restoration}

The chiral symmetry restoration is studied \cite{asy}, \cite{ps}
by introducing flavor D8 and 
anti-D8 branes  placed at two different points of the compact coordinate 
$\tau$, which has the periodicity $\tau\sim\tau+2\pi$. The gauge symmetry 
on these 
branes is interpreted following Sakai-Sugimoto as the chiral symmetry 
$U(N_f)\times U(N_f)$. In the 
zero temperature both the branes and anti-branes get joined together so as 
to break this symmetry to its diagonal subgroup whereas in the finite
 temperature case they get separated and finally end on the horizon. This 
form the field theory point of view is interpreted as the chiral symmetry
restoration. 

Here we would like to see how do these global symmetry $U(N_f)\times U(N_f)$
 breaks to its diagonal subgroup. Probably,  its difficult to define chirality
in a theory that admits non-relativistic symmetry, so instead we shall 
interpret the breakdown of the symmetry as a breakdown of a global symmetry
and chiral symmetry means we are talking about the global symmetry 
$U(N_f)\times U(N_f)$. 

In order to proceed and do the calculation in this setting let us recall 
that the dynamics of these flavor branes are described by the DBI and CS 
actions, which   are
\be\label{dbi}
S_{DBI}=-T_8\int d^9\sigma e^{-\Phi}
\sqrt{-det([g+B]_{ab}+ F_{ab})}
\ee 
 
and \be
S_{CS}=\mu_8 \int \sum_n [C^{(n)}\w e^{B}]\w e^{\lambda F},
\ee

where [~~] denotes pull-back of the space time fields onto the world volume 
of the 
brane. In the case of interest, it is trivial to see that the CS action is not
 going to contribute. Hence, the only contribution comes from the DBI part.
Also, in the action we have included gauge field in the action, whose 
interpretation from the field theory point of view is that turning on the 
chemical potential for the 
baryon number, for non-trivial electric field, $F_{0r}=-\p_r {{\cal A}}_0$, 
\cite{ht}.

The induced metric on the flavor brane, by assuming that we are only exciting 
the s-wave part of the  mode i.e. taking the embedding to be a function of 
only the radial 
direction, for the most general solution after step 6 i.e. 
the eqs (\ref{gen_metric}), is

\bea\label{gen_ind_metric} ds^2&=&\f{A}{K}(1+\beta^2 DB sin^2\theta)dt^2+\f{B}{K}(1+\beta^2 DA sin^2\theta)dy^2+2 \f{ABD\beta^2 sin^2\theta}{K}dtdy\nn&+&\delta_{11}(r) dx^1 dx^1+\delta_{22}(r) dx^2dx^2+[\delta_{33}(r) \tau'(r)^2 +C] dr^2+D d\theta^2+D sin^2\theta d\varphi^2\nn&+&\f{D}{K} sin^2\theta sin^2\varphi(1+\beta^2 D sin^2\theta cos^2\varphi (A+B))d\psi^2\nn&+&\f{D}{K} sin^2\theta cos^2\varphi (1+\beta^2 D sin^2\theta sin^2\varphi(A+B))d\chi^2\nn&-&2 \f{\beta^2 D^2 sin^4\theta sin^2\varphi cos^2\varphi(A+B)}{K}d\chi d\psi,
\nn
\eea 
where $\tau'(r)=\f{d\tau}{dr}$, whose $\beta \rightarrow 0$ limit will give the
the induced metric on the flavor brane of the geometry eq(\ref{d4_sol}) and
taking the functions $A,~~ B,~~ C,~~D,~~ \delta_{ii}$ and K appropriately 
can give us the 
induced metric for a black hole or at zero temperature, i.e. the choice of 
eq(\ref{bh_functions}) and eq(\ref{ext_functions}), respectively.

The square root of the  determinant of the relevant quantity 
$\left([g+B]_{ab}+F_{ab}\right)$ with the 
induced geometry written in eq(\ref{gen_ind_metric})
\be\label{dbi_action}
\sqrt{-det([g+B]_{ab}+F_{ab})}=\f{D^2 sin^3\theta sin~2\varphi
\sqrt{B\delta_{11}\delta_{22}
[{\cal A}_0'^2+A(C+\delta_{33}\tau'^2)]}}{2\sqrt{1+\beta^2D(A+B)sin^2\theta}},
\ee
where we have only turned on the electric flux associated to the U(1) field 
strength i.e. $F_{r0}$. The dilaton
\be
e^{-\Phi}=e^{-\Phi_0}\sqrt{1+\beta^2D(A+B)sin^2\theta}.
\ee

It means the action of the flavor brane 
\bea\label{d8_action}
S_{DBI}&=&-T_8\int e^{-\Phi}\sqrt{-det([g+B]_{ab}+F_{ab})}=\nn&-&T_8\int e^{-\Phi_0}\f{D^2}{2} sin^3\theta 
sin~2\varphi\sqrt{B\delta_{11}\delta_{22}[{\cal A}_0'^2+A(C+\delta_{33}\tau'^2)]}
\eea
which is independent of the parameter $\beta$. It just follows trivially 
that the action of the flavor branes before and after the Null Melvin Twists
are the same. This can be confirmed by calculating the action of the flavor 
branes from the 10 dimensional geometry eq(\ref{d4_sol}) 
 with an electric 
field, $F_{r0}$, turned on the world volume of the flavor brane, for the 
functions, as written in eq(\ref{bh_functions}). Similarly, evaluating 
eq(\ref{d8_action})  using the functions as written in 
eq(\ref{ext_functions})   gives the same answer as that using  
the 10 dimensional geometry eq(\ref{d4_sol}) and eq(\ref{d8_action}).

The consequence of all this is that the chiral symmetry breaking
and restoration curve remains unchanged with respect to Null Melvin Twist. 
This can be seen very easily  by taking the difference
of the flavor branes evaluated on the non-zero and zero temperature 
backgrounds. Its because,  we saw that the actions 
for the flavor branes remain same under Null Melvin Twist. As a result, the
distance of separation between the quarks, L, remain same and hence the
curve that describes the chiral symmetry  restoration curve $LT=c$, 
remains same, where c is a constant and whose value is less than one.

\section{Confinement and De-confinement transition}
The confinement and de-confinement transition is a property of the bulk 
solution \cite{ew} and 
can be studied by computing the on-shell action which by AdS/CFT is related to 
the free energy of the system. The action of the D4 brane with the non-trivial
 degrees of freedom such as  metric, dilaton, 2-from and 4-form antisymmetric 
fields, $g_{MN},~\Phi,~B_2,~F_4$, respectively,  is 
\bea
S&=&\f{1}{2\kappa^2}\int d^{10}x\sqrt{-g}[R_E-\f{1}{2}g^{MN}\p_M\Phi\p_N\Phi-
\f{g_s e^{-\Phi}}{12}H_{MNP}H^{MNP}-\nn&&
\f{g^{3/2}_s e^{\Phi/2}}{48}F_{MNPQ}F^{MNPQ}]  
\eea
The trace of the equation motion of  metric give
\be
R_E-\f{1}{2}g^{MN}\p_M\Phi\p_N\Phi-
\f{g_s e^{-\Phi}}{24}H_{MNP}H^{MNP}-\f{g^{3/2}_s e^{\Phi/2}}{192}F_{MNPQ}F^{MNPQ}=0
\ee

Now we shall eliminate $H_{MNP}H^{MNP}$ and write down the action in terms of 
other fields
\be
\label{d4_action}
S=-\f{1}{2\kappa^2}\int d^{10}x\sqrt{-g}[R_E-\f{1}{2}g^{MN}\p_M\Phi\p_N\Phi+
\f{g^{3/2}_s e^{\Phi/2}}{96}F_{MNPQ}F^{MNPQ}] 
\ee

In order to compute the on shell action we need to know each  terms that 
appear in the action. 
Various quantities  of  black hole solution eq(\ref{solI_e}) are
\bea\label{bh_ex_e}
\sqrt{-g}&=&r^{15/8}R^{15/8}sin^3\theta sin\varphi cos\varphi
(r+\beta^2 r^3_Tsin^2\theta)^{1/4}, \nn
R_E&=&\f{9(5r^3-r^3_T)}{32 r^{25/8}R^{15/8}}+\nn&&\beta^2r^3_T
\f{131r^3-71 r^3_T+(381 r^3+71 r^3_T)cos~2\theta}{256 r^{33/8}R^{15/8}}+
{\cal O}(\beta^4),\nn
g^{MN}\p_M\Phi\p_N\Phi&=&\f{9(r^3-r^3_T)}{16 r^{25/8}R^{15/8}}+\beta^2r^3_T
\f{39 (r^3-r^3_T)sin^2\theta}{64 r^{33/8}R^{15/8}}+{\cal O}(\beta^4),\nn
e^{\Phi/2}F_{MNPQ}F^{MNPQ}&=&\f{f^2_0\sqrt{g_s}r^{1/8}}{R^{63/8}
(r+\beta^2 r^3_T sin^2\theta)^{1/4}}
\eea

Adding all these terms with the appropriate coefficients that appear in 
 action eq(\ref{d4_action}) 
\bea
S_{bh}&=&-\f{1}{2\kappa^2}\int^{\beta_H}_{0} dt\int [dy dx dz]
\int^{R_{\star}}_{r_T}dr\int^{\pi}_{0}d\theta\int^{\pi/2}_{0}d\varphi\int^{2\pi}_{0}d\psi\int^{2\pi}_{0}d\chi\nn&&\int^{2\pi {\tilde R}_{\tau}}_{0}d\tau
  sin^3\theta sin\varphi cos\varphi
\bigg[\f{r^2}{96 R^6}(f^2_0g^2_s+108R^6)+\nn&&\f{\beta^2r^3_T}{8 r^2}\bigg(4r^3-r^3_T+(12r^3+r^3_T)cos~2\theta\bigg)+{\cal O}(\beta^4)\bigg],
\eea
where $\beta_H=\f{1}{T_H}$ and $2\pi {\tilde R}_{\tau}$ is the periodicity
of the $\tau$ circle and the range of integration for the radial coordinate is
from the horizon to a large value $R_{\star}$. 
For the extremal solution eq(\ref{solII_e}), various expressions are
\bea\label{ex_ex_e}
\sqrt{-g}&=&r^{17/8}R^{15/8}sin^3\theta sin\varphi cos\varphi,~~
R_E=\f{9(5r^3-r^3_0)}{32 r^{25/8}R^{15/8}},\nn
g^{MN}\p_M\Phi\p_N\Phi&=&\f{9(r^3-r^3_0)}{16 r^{25/8}R^{15/8}},~~
e^{\Phi/2}F_{MNPQ}F^{MNPQ}=\f{f^2_0\sqrt{g_s}}{r^{1/8}R^{63/8}}
\eea

It is easy to see that in the $\beta\rightarrow 0$ limit, eq(\ref{bh_ex_e})
goes over to eq(\ref{ex_ex_e}). Note that the Ricci-scalar of eq(\ref{solII_e})
do not depend on the parameter $\beta$, even though the solution do. 
Now, the action for the extremal case
\bea
S_{ex}&=&-\f{1}{2\kappa^2}\int [dy dx dz]
\int^{2\pi  R_{\tau}}_{0}d\tau\int^{R_{\star}}_{r_0}dr\int^{\pi}_{0}d\theta\int^{\pi/2}_{0}d\varphi\int^{2\pi}_{0}d\psi\int^{2\pi}_{0}d\chi\int^{{\tilde \beta}}_{0} dt\nn&&
  sin^3\theta sin\varphi cos\varphi
\bigg[\f{r^2}{96 R^6}(f^2_0g^2_s+108R^6)\bigg],
\eea

where ${\tilde \beta}$ is the periodicity of the Euclidean time circle and 
$2\pi  R_{\tau}$ is the periodicity of the $\tau$ circle in the extremal case.

The periodicity  ${\tilde \beta}$ is not arbitrary but is computed by 
taking the circumference of the Euclidean time circles be same for both
the black hole and the extremal case,  
for a large value of the radial coordinate \cite{ew} 
\be
\int^{{\tilde \beta}}_0 dt\sqrt{g_{tt}}|_{r=R_{\star}}=\int^{ \beta_H}_0 dt\sqrt{g_{tt}}|_{r=R_{\star}}
\ee
and is 
related to the periodicity $\beta_H$ 
of the black hole as
\be
{\tilde\beta}=\beta_H\bigg[1-(\f{r_T}{R_{\star}})^3\bigg]^{1/2}\bigg[1+\f{r^3_T}{R_{\star}}\beta^2 sin^2\theta\bigg]^{-3/8}
\ee 
and similarly for the compact $x^4=\tau$ circle
\be
{\tilde R}_{\tau}=R_{\tau}\f{\sqrt{1-(r_0/R_{\star})^3}}{[1+\f{r^3_T}{R_{\star}}\beta^2 sin^2\theta]^{1/8}}
\ee

Now doing the appropriate integrals and keeping terms to quadratic order 
in $\beta$ results
\bea
S_{bh}&=&-\f{\beta_H}{2\kappa^2} V_3 2\pi^2 2\pi R_{\tau}\Bigg[
\f{f^2_0 g^2_s+108 R^6}{216 R^6}R^3_{\star} -\beta^2 
\f{f^2_0 g^2_s+684 R^6}{2160 R^6}r^3_TR^2_{\star}\nn&-&
\f{f^2_0 g^2_s+108 R^6}{432 R^6}(r^3_0+2 r^3_T)+
{\cal O}\bigg(\beta^4,\f{1}{R_{\star}} \bigg)\Bigg]\nn
S_{ex}&=&-\f{\beta_H}{2\kappa^2} V_3 2\pi^2 2\pi R_{\tau}\Bigg[
\f{f^2_0 g^2_s+108 R^6}{216 R^6}R^3_{\star} -\beta^2 
\f{f^2_0 g^2_s+108 R^6}{720 R^6}r^3_TR^2_{\star}\nn&-&
\f{f^2_0 g^2_s+108 R^6}{432 R^6}(2 r^3_0+ r^3_T)+
{\cal O}\bigg(\beta^4,\f{1}{R_{\star}} \bigg)\Bigg]
\eea

The difference of the actions 
\bea\label{delta_s} 
\Delta S=S_{bh}-S_{ex}&=&-\f{\beta_H}{2\kappa^2} V_3 2\pi^2 2\pi R_{\tau}\Bigg[
(5 f^2_0 g^2_s+540 R^6)(r^3_0-r^3_T)\nn&+&2 \beta^2 R^2_{\star} r^3_T
\f{(f^2_0 g^2_s-108 R^6)}{2160 R^6}+{\cal O}\bigg(\beta^4,\f{1}{R_{\star}} \bigg)\Bigg]
\eea

In the second line of eq(\ref{delta_s}) we see the appearance of a UV divergent
term. This arises because we have not  added the proper counter term to 
regulate  the integrals. What we shall do is to drop this term and consider
only the finite term. Of course, it is important and interesting to 
understand the structure of the necessary counter terms.

From the finite term of the difference of the black hole and confining actions 
we see that it is proportional to $r^3_0-r^3_T$ and is independent of the
parameter $\beta$.
 Recalling the periodicity
as written in eq(\ref{periodicities}) we finally get the difference as
\be 
\Delta S=-\f{V_3}{2\kappa^2}4\pi^3  R_{\tau}\beta_H (5 f^2_0 g^2_s+540 R^6)
\bigg(\f{16\pi^2 R^3}{9}\bigg)^3 
\Bigg[\f{1}{\beta^6_H}-\f{1}{(2\pi R_{\tau})^6} \Bigg]
\ee

To see that the action go as $\Delta S\sim N^2_C (g_s N_c)$ in the large 
$N_c$ and large $g_s N_c$ limit, recall that\footnote{In the 
$2\pi l_s=1$ unit.} 
$2\kappa^2\sim g^2_s$,~~
$f^2_0 g^2_s\sim~ R^6$,~~ $\beta_H R_{\tau}\sim R^3$ and $R^3\sim g_s N_c$means 
\be
\Delta S\sim \f{1}{g^2_s} g^2_s N^2_c \bigg(\f{R^9 \beta_H R_{\tau}}{\beta^6_H}\bigg)\sim N^2_C (g_s N_c)
\ee

In the limit $R_{\tau} T_H < \f{1}{2\pi}$, then the extremal configuration has 
got lower free energy and is preferred and if $R_{\tau} T_H > \f{1}{2\pi}$
then the black hole phase is preferred. The curve that describes the confining
 de-confining phase transition  is exactly the same as in the 
$\beta\rightarrow 0$ limit.

\section{Conclusion}

We have analyzed different aspects of Null Melvin twist to  D4 brane solution
with the asymptotic to that of a  plane wave. The resulting background 
at zero temperature shows confinement and at finite temperature becomes a black
hole. The result of this study is showing some interesting properties
 that is the 
curve that describes the confining de confining phase 
transition remains unchanged
so also all the periodicity at zero and non-temperature. Upon inclusion of the
flavors via a bunch of coincident
D8 and anti-D8 branes shows that the curve that 
describes the chiral symmetry breaking and restoration remains unchanged.

The vev associated to the operator which transforms bi-fundamentally under the 
color gauge group, for the breakdown of the chiral symmetry i.e. 
the chiral condensate can be calculated using the top-down prescription 
\cite{ak} and \cite{hhly} which is given by computing the expectation 
value of the open 
Wilson line that is stretched between the D8 and anti-D8 branes and the world
sheet of the Wilson line that  is extended along  $r$ and  $\tau$ directions. 
 It is interesting  to know  that the $(r,\tau)$ part of the metric written
in eq(\ref{solII_e}), do not depends on the parameter $\beta$, 
which means the chiral 
condensate  also remains unchanged under Null Melvin Twist. 

It has certainly become very interesting to understand more about the solution
presented in \cite{klm}. Unlike the solutions \cite{son}, \cite{bm},\cite{hrr}
and \cite{abm}, this solution has got an intrinsic exponent z, in the 
sense that  all the coordinate invariant quantities depends on this 
exponent in a very non-trivial way and more importantly, the
 solution is not of the plane wave type.

{\bf Acknowledgment}: It is a pleasure to thank Ofer Aharony, 
Mukund Rangamani, Simon Ross and Cobi Sonnenschein for useful correspondences.
I would like to thank
  Rabin Banerjee for the kind invitation to
visit  S N Bose centre for basic sciences, Kolkata, and the institute 
for providing a warm hospitality, where a part of the work is done.


\begin{thebibliography}{99}

\bibitem{my}T. Maskawa and K. Yamawaki, {\it Rho problem of $P_+=0$ mode in 
the null plane field theory and Dirac's method of quantization}, 
Prog. Theor. Phys. {\bf 56} (1976) 270.
\bibitem{abs} O. Aharony, M. Berkooz and N. Seiberg, {\it Light cone 
description of (2,0) superconformal theories in six dimensions}, Adv. Theor. Math. Phys. {\bf 2} (1998) 119, hep-th/9712117.
\bibitem{mmt} J. Maldacena, D. Martelli and Y. Tachikawa, {\it Comments on
string theory backgrounds with non-relativistic conformal symmetry}, arXiv:0807.1100 [hep-th].
\bibitem{ghhlr}E. G. Gimon, A. Hashimoto, V. E. Hubeny, O. Lunin and 
M. Rangamani, {\it Black strings in asymptotically plane wave geometries}, JHEP {\bf 08} (2003) 035, hep-th/0306131.
\bibitem{ag}M. Alishahiha and O. J. Ganor, {\it Twisted backgrounds, pp-waves
and non-local field theories}, JHEP {\bf 03} (2003) 006, hep-th/0301080.

\bibitem{hrr}C. P. Herzog, M. Rangamani and S. F. Ross, 
{\it Heating up Galilean holography}, arXiv:0807.1099[hep-th].
\bibitem{abm}A. Adams, K. Balasubramanian and J. McGreevy, {\it Hot spacetimes 
for cold atoms}, arXiv:0807.1111[hep-th].
\bibitem{son}D. T. Son, {\it Towards an Ads/cold atoms correspondence: a 
geometric realization of the Schr$\ddot{o}$dinger symmetry}, arXiv:0804.3972[hep-th].
\bibitem{bm}K. Balasubramanian and J. McGreevy, {\it Gravity duals for 
non-relativistic CFTs}, arXiv:0804.4053[hep-th].
\bibitem{wg} W. D. Goldberger, {\it AdS/CFT duality for non-relativistic 
field theory}, arXiv:0806.22867 [hep-th].
\bibitem{bf}J. L. Barbon and C.A. Fuertes, {\it On the spectrum of 
non-relativistic AdS/CFT}, arXiv:0806.3244 [hep-th].
\bibitem{ch}C. R. Hagen, {\it Scale and conformal transformations in galilean-
covariant field theory}, Phys. Rev. {\bf D5} (1972)377-388.
\bibitem{msw} T. Mehen, I. W. Stewart and M. B. wise, {\it Conformal invariance 
for non-relativistic field theory}, Phys. Lett. {\bf B474} (2000), 145-152, hep-th/9910025.
\bibitem{ns} Y. Nishida and D. T. Son, {\it Nonrelativistic conformal field theories}, Phys. Rev.{\bf D76} (2007)086004, arXiv:0706.3746.
\bibitem{wy}W. Y. Wen, {\it AdS/NRCFT for the (super) calogero model}, 
arXiv:0807.0633.
\bibitem{thron}  C. B. Thorn, {\it Asymptotic freedom in the infinite-momentum frame}, Phys. Rev. {\bf D 20},(1979) 1934; {\it Fock-space description of the $1/N_c$ expansion of quantum chromodynamics}, Phys. Rev. {\bf D 20},(1979) 1435; {\it Quark confinement in the infinite-momentum frame},  Phys. Rev. {\bf D 19},(1979) 639; O. Bergman and C. B. Thorn, {\it Super-Galilei invariant field theories in 2+1 dimensions}, Phys. Rev. {\bf D 52}, (1995) 5997.
\bibitem{klm} S. Kachru, X. Liu and M. Mulligan, {\it Gravity duals of 
Lifshitz-like fixed points}, arXiv:0808.1725[hep-th].
\bibitem{ssp}S. S. Pal, {\it Towards Gravity solutions of AdS/CMT}, 
arXiv:0808.3232[hep-th].
\bibitem{nps} R. R. Nayak, K. L. Panigrahi and S. Siwach, {\it Brane solutions
with/without rotation in pp-wave spacetime}, Nucl. Phys. {\bf B 698}, (2004)
149-162, hep-th/0405124.
\bibitem{bfhp} M. Blau, J. Figueroa-O'Farrill, C. Hull and G. Papadopoulos, {\it A new maximally supersymmetric background of IIB superstring theory}, JHEP {\bf 01} (2002) 047, hep-th/0110242.
\bibitem{bmn} D. Berenstein. J. Maldacena and H. Nastase, {\it Strings in flat
space and pp waves from ${\cal N}=4$ super Yang Mills}, JHEP 0204 (2002) 013,
hep-th/0202021.

\bibitem{mr} V. E. Hubeny and M. Rangamani, {\it Generating asymptotically 
plane wave spacetimes}, JHEP, 0301 (2003) 031; {\it Horizons and plane waves: A
review}, Mod. Phys. Lett. {\bf A 18}, (2003) 2699-2712.
\bibitem{hrr1} V. E. Hubeny, M. Rangamani and S. F. Ross, {\it Causal 
inheritance in plane wave quotients}, Phys. Rev. {\bf D 69} (2004) 024007, 
hep-th/0307257.
\bibitem{st} K. Skenderis and M. Taylor, {\it Branes in AdS and pp-wave spacetimes}, JHEP 0206 (2002) 025, hep-th/0204054; {\it An overview of branes in the 
plane wave background}, Class. Quant. Grav. {\bf 20} (2003), hep-th/0301221. 
\bibitem{pal1}S. S. Pal, {\it Solution to worldvolume action of D3 brane in 
PP-wave background}, Mod. Phys. Lett. {\bf A 17}, (2002) 1735-1744.
\bibitem{as} M. Ali-Akbari and M. M. Sheikh-Jabbari, {\it Electrified BPS 
giants: BPS configurations on giant gravitons with static electric field}, JHEP
0710 (2007) 043, 0708.2058[hep-th].
\bibitem{kk} A. Kumar and H. K. Kunduri, {\it Gravitational wave solutions in
string and M-theory AdS backgrounds}, Phys. Rev. {\bf D 70} (2004) 104006, 
hep-th/0405261.
\bibitem{pm}P. Mukhopadhyay, {\it Tachyon condensation and non-bps D-branes in
Ramond-Ramond plane wave background}, hep-th/0611138.
\bibitem{tm}T. Mattik, {\it Branes in plane wave background with gauge
field condensates}, JHEP, 0506 (2005) 041, hep-th/0501088.
\bibitem{dg}S. R. Das and C. Gomez, {\it Realizations of conformal and 
Heisenberg algebras in PP wave and CFT correspondence}, JHEP 0207 (2003) 016,hep-th/0206062.
\bibitem{zs} L. A. Pando Zayas and J. Sonnenschein, {\it On Penrose limits and
gauge theories}, JHEP 0205 (2002) 010.

 




\bibitem{ew}
  E.~Witten,
  {\it Anti-de Sitter space, thermal phase transition, and 
  confinement in  gauge
  theories},
  Adv.\ Theor.\ Math.\ Phys.\  {\bf 2}, 505 (1998);
  hep-th/9803131.  

\bibitem{kk}
 A.~Karch and A.~Katz,
 {\it Adding flavor to AdS/CFT},
  JHEP {\bf 0206}, 043 (2002);
  hep-th/0205236.

\bibitem{ekss} J. Erlich, E. Katz, D. T. Son and M. A. Stephanov, {\it QCD and
a holographic model of hadrons}, Phys. Rev. Lett. {\bf 95} (2005) 261602, hep-th/0501218.

\bibitem{gk} U. Gursoy and E. Kiritsis, {\it Exploring improved holographic theories for QCD: Part I}, 0707.1324[hep-th].

\bibitem{gkn} U. Gursoy,  E. Kiritsis and F. Nitti, {\it Exploring improved holographic theories for QCD: Part II}, 0707.1349[hep-th].
    
\bibitem{lm}
  O.~Lunin and J.~Maldacena,
  {\it Deforming field theories with U(1) $\times$ U(1) global symmetry and their gravity 
duals}
  JHEP {\bf 0505}, 033 (2005)
  [arXiv:hep-th/0502086].

\bibitem{ssp1}S.~S.~Pal,
  {\it beta-deformations, potentials and KK modes,}
  Phys.\ Rev.\ D {\bf 72} (2005) 065006
  [arXiv:hep-th/0505257].

 
\bibitem{ss}
  T.~Sakai and S.~Sugimoto,
  {\it Low energy hadron physics in holographic QCD},
  Prog.\ Theor.\ Phys.\  {\bf 113}, 843 (2005);
  hep-th/0412141.
  
 
\bibitem{imsy}
  N.~Itzhaki, J.~M.~Maldacena, J.~Sonnenschein and S.~Yankielowicz,
  {\it Supergravity and the large N limit of theories with sixteen
  supercharges},
  Phys.\ Rev.\ D {\bf 58}, 046004 (1998)
  [arXiv:hep-th/9802042].
 
\bibitem{ahjk}
  E.~Antonyan, J.~A.~Harvey, S.~Jensen and D.~Kutasov,
  {\it NJL and QCD from string theory},
  arXiv:hep-th/0604017.
 

 

 
\bibitem{asy} O. Aharony, J. Sonnenschein and S. Yankielowciz, {\it A 
holographic model of deconfinement and chiral symmetry breaking}, 
arXiv:hep-th/0604161.
\bibitem{ps}
  A.~Parnachev and D.~A.~Sahakyan,
  {\it Chiral phase transition from string theory},
  arXiv:hep-th/0604173.
 

\bibitem{ahk}E.~Antonyan, J.~A.~Harvey, and D.~Kutasov, {\it The 
 Gross-Neveu model from string theory},  arXiv:hep-th/0608149; {\it Chiral 
 symmetry breaking from intersecting D-branes},  arXiv:hep-th/0608177.
\bibitem{ht} N. Horigome and Y. Tanii, {\it Holographic chiral phase 
transition with chemical potential}, JHEP 0701 (2007) 072, hep-th/0608198.

\bibitem{gp} D.~Gepner and S.~S.~Pal, {\it Chrial symmetry breaking and 
restoration from holography}, hep-th/0608229.

\bibitem{ak} O. Aharony and D. Kutasov, {\it Holographic duals of long open 
strings}, arXiv:0803.3547 [hep-th].
\bibitem{hhly} K. Hashimoto, T. Hirayama, F. -L. Lin and H. -U. Yee, {\it Quark mass deformation of holographic massless QCD}, arXiv:0803.4192[hep-th].
\bibitem{ckp}R. Casero, E. Kiritsis and A. Paredes, {\it Chiral symmetry 
breaking as open string tachyon condensation}, Nucl. Phys. {\bf B 787}, (2007) 98, hep-th/0702155.
\bibitem{bss} O. Bergman, S. Seki and J. Sonnenschein, {\it Quark mass and 
condensation in HQCD}, JHEP {\bf 0712} (2007) 037, arXiv:0708.2839[hep-th].
\bibitem{dn} A. Dhar and P. Nag, {\it Sakai-Sugimoto model, tachyon condensation and chiral symmetry breaking}, JHEP {\bf 0801}, (2008) 055, 
arXiv:0708.3233[hep-th];{\it 
Tachyon condensation and quark mass in modified Sakai-Sugimoto model}, 
arXiv:0804.4807[hep-th].
\bibitem{nms} R. McNees, R. C. Myers and A. Sinha, {\it On quark masses in holographic QCD}, arXiv:0807.5127[hep-th].



\end{thebibliography}
\end{document}